\documentstyle[11pt]{article} 
\input epsf

\begin{document} 

\title{Random Walk with a Hop-Over Site:\\ 
A Novel Approach to Tagged Diffusion and Its Applications} 
\author{R.K.P.Zia and Z.Toroczkai \\ 
Physics Department and Center 
for Stochastic Processes in\\ 
Science and Engineering, Virginia
 Polytechnic Institute and State University, 
\\ 
Blacksburg, V.A. 24061-0435, USA}  

\maketitle 
 
\begin{abstract} 
We study, on a $d$ dimensional 
hypercubic lattice, a random walk which is 
homogeneous except for one site.
 Instead of visiting this site, the walker 
hops over it with arbitrary rates.
 The probability distribution of this walk 
and the statistics associated with 
the hop-overs are found exactly. This 
analysis provides a simple approach 
to the problem of tagged diffusion, 
i.e., the movements of a tracer 
particle due to the diffusion of a vacancy. 
Applications to vacancy mediated 
disordering are given through two examples. 
\end{abstract} 

\vfill

\noindent $\overline{\mbox{{\bf KEY WORDS}:
 random walk, lattice defects, tagged diffusion, 
 vacancy-mechanism}}$
 
\newpage 
 
\section{Introduction} 
 
The venerable problem of a random 
walk on a regular $d$-dimensional 
hypercubic, infinite lattice continues
 to generate considerable interest  
\cite{H}-\cite{Itzykson}, from both novel quantities 
associated with the simple walk and new 
variations of the walk itself. In the 
simplest case (which is called P\'olya walk 
\cite{Polya}), 
the walker moves to one 
of the nearest neighbor sites with 
probability $1/2d$ at each time step. 
One of the variations involves a ``taboo''
site, to which the walker may never 
visit. The question that is usually asked refers
to the probability \cite{Chung}, \cite{H} (``taboo 
probability'') of the walker
visiting a certain site on the $n$-th step without
having visited the taboo site on any of the 
erlier steps $1,2,...,n-1$. An alternative way to think
about the taboo site is to consider it as an irreversible
trap, so that the above question translates into
a question on the survival probability .
A slightly different situation is when
the walker arriving at a 
neighboring site of the taboo, is let to remain 
stationary, with probability 
$1/2d$, instead of moving into the 
forbidden site. All these variations belong to the
chapter of lattice walks with ``defective'' sites in the
theory of lattice walks \cite{H}.  
In this paper, we 
investigate a further case of walks with a defective
site, namely, the 
walker {\em hopping over} this special 
site. Since it is no longer 
``taboo'' (in the above sense), it will be referred to as 
the ``hop-over'' site. We write the 
Master equation for such a walk, with 
$2d$ arbitrary hop-over rates. Thanks 
to translational invariance, we may choose
 to locate the hop-over site at 
the origin. In Fourier  space, we obtain 
a closed expression, in terms of the 
inverse of a matrix, for the probability
 distribution generating function, 
given a particular initial position ${\bf s_0}$. 
 
Similar to the derivation of ``taboo 
probabilities'', the straightforward
way to find the hop-over probabilities 
would rely on a study of return
probabilities and first passage times. 
In this approach, probabilities of
returning to the neighborhood of the 
origin {\em without} hopping over
it are exploited in infinite sums over 
individual hop-over attempts. A
similar method was used for solving 
the related problem of tagged diffusion
\cite{ZT,BH}. Our present approach is different,
considerably simpler and more
compact in terms of formulas. Instead of
infinite sums, only the inverse of
a certain $2d\times 2d$ matrix needs to be
computed. In $d=2$, where this walk displays
the most interesting characteristics, the
inversion is quite simple.

We are also able to keep track of both 
the direction and the frequency of 
the hop-over attempts. As a result, an 
interesting application is the 
diffusive behavior of a tagged particle
 \cite{ZT,BH}. Such a tag may also be 
thought of as a ``passive walker'', 
namely, a particle which remains inert 
until it is forced to exchange 
places with a normal, ``active'' random 
walker. In metals or alloys, an 
impurity atom can play the role of a tag, 
while a vacancy diffuses as a typical
 walker \cite{Sholl}-\cite{Yuri}.
 We will discuss how to map the 
hop-over problem to the tagged walk.
 Finally, from the frequency of hop-over 
attempts, we can compute the distribution
 of ``hits'' received by the tag. 
We present two related applications: the first 
studies  the rate of disordering
 of the surface of an A-B alloy, 
with a certain type of interactions, due
 to the wandering of a surface vacancy,
 while the second investigates 
 the asymptotic time evolution 
of the disorder caused by a single Brownian
vacancy in an initially completely phase seggregated, 
two-species system.   
 
This paper is organized as follows. In
 the next section, the model and the 
Master equation are carefully defined 
(Sections 2.1 and 2.2) and the 
explicit solution is presented allowing 
for the extraction of the statistics 
on the hop-over events (Section 2.3). 
Section 3 presents the relationship to 
the tagged diffusion problem with its 
complete solution through the 
techniques of Section 2., along with two 
associated applications presented 
in Sections 3.2 and 3.3. The last Section is
devoted to a summary and 
possible generalizations. 
 
\section{Random Walk with a Hop-Over Site} 
 
For completeness, we devote the 
first subsection to definitions, notations, 
and some well known properties
 of a random walk on a lattice. 
 
\subsection{The simple random walk} 
 
On a infinite $d$-dimensional hypercubic
lattice, the sites are labeled by ${\bf s}$ 
while the set of 2d lattice vectors is 
denoted by $\{{\bf a}\}$. A walker, performing a 
pure random walk of P\'{o}lya type \cite{H}, 
\cite{Polya} moves 
from a site to one of the $2d$ nearest 
neighbor sites at each time step 
(i.e., from ${\bf s}$ to ${\bf s+a}$) 
with probability $p$. If $2dp<1$, the 
walker remains stationary with probability $1-2dp$. 
Given an initial 
position ${\bf s_0}$ we are interested 
in the probability distribution of 
finding the walker after $n$ steps at 
site ${\bf s}$ : 
$P_n^F({\bf s}|{\bf s_0})$ . Here, the superscript 
$F$ reminds us that this is the 
distribution for ``free'' diffusion. 
%In order to ease the notations, in the 
%following we shall suppress variable 
%${\bf s_0}$. 
The equation governing 
this walk is  
\begin{equation} 
P_{n+1}^F({\bf s}|{\bf s_0})-P_n^F({\bf s}|{\bf s_0})=
p\sum_{\{{\bf a}\}}
\left[ P_n^F({\bf s+a}|{\bf s_0})-
P_n^F({\bf s}|{\bf s_0})\right] \quad ,  \label{freeME} 
\end{equation} 
with the initial condition:  
\begin{equation} 
P_0^F({\bf s}|{\bf s_0})=\delta _{{\bf s},
{\bf s_0}}\quad .  \label{freeinit} 
\end{equation} 
Due to translational invariance, the 
solution can be obtained simply in 
terms of the generating function in Fourier space:  
\begin{eqnarray} 
P({\bf s}|{\bf s_0};\xi ) &\equiv &\sum_{n=0}^\infty 
P_n({\bf s}|{\bf s_0})\xi ^n\quad 
,\;\;\;|\xi|<1,  \label{gendefP} \\ 
\tilde{P}({\bf k};\xi ) &\equiv 
&\sum_{{\bf s}}P({\bf s}|{\bf s_0};\xi )
e^{i{\bf k\cdot s}}\quad .  \label{gendefF} 
\end{eqnarray} 
Applying (\ref{gendefP}) to ${\bf s=0}$ in 
(\ref{freeME}), we obtain a useful identity  
\begin{equation} 
\xi p\sum_{\{{\bf a}\}}P^F({\bf a}|{\bf s_0};\xi )=
\left[ P^F({\bf 0}|{\bf s_0};\xi )-\delta _{ 
{\bf 0,s}_{{\bf 0}}}\right] +\xi 
(2dp-1) P^F({\bf 0}|{\bf s_0};\xi )\;. 
\label{sumPbxi} 
\end{equation} 
 
Continuuing, we substitute (\ref{freeME}) 
into (\ref{gendefP}) and (\ref 
{gendefF}) to arrive at  
\[ 
\frac 1\xi \left[ \tilde{P}^F({\bf k};\xi )
-e^{i{\bf k}{\bf s_0}}\right] - 
\tilde{P}^F({\bf k};\xi )=
-p\sum_{\{{\bf a}\}}\left[ 1-\cos ({\bf k\cdot } 
{\bf a})\right] \tilde{P}^F({\bf k};\xi )  
\] 
The solution is trivial:  
\begin{equation} 
\tilde{P}^F({\bf k};\xi )=
G({\bf k},\xi )e^{i{\bf k}{\bf s_0}} 
\label{freeRW} 
\end{equation} 
where  
\begin{equation} 
G({\bf k},\xi )\equiv \left\{ 1-\xi +
\xi p\sum_{\{{\bf a}\}}\left[ 1-\cos( 
{\bf k\cdot }{\bf a})\right] \right\} ^{-1}  \label{G} 
\end{equation} 
is the well known propagator for free diffusion. 
 
The inverses to (\ref{gendefF}, \ref{gendefP}) are given by  
\begin{equation} 
P({\bf s}|{\bf s_0};\xi )=\int_{{\bf k}}
e^{-i{\bf k\cdot s}}\tilde{P}({\bf k};\xi ) 
\label{invertS} 
\end{equation} 
and  
\begin{equation} 
P_n({\bf s}|{\bf s_0})=\oint_\xi 
\xi^{-n}P({\bf s}|{\bf s_0};\xi )\quad ,  \label{invertN} 
\end{equation} 
where  
\[ 
\int_{{\bf k}}\equiv \int\limits_{-\pi }^\pi
 \frac{d^dk}{(2\pi )^d}\quad % 
\mbox{and}\quad \oint_\xi \equiv \frac{1}{2\pi i}
\oint\limits_{{\cal C}}\frac{d\xi }\xi  
\] 
and ${\cal C}$ is a suitable (counterclockwise)
 contour around $\xi =0$. 
Thus, the solution to the simple 
random walk can be written as  
\begin{equation} 
P^F({\bf s}|{\bf s_0};\xi )=
\int_{{\bf k}}e^{-i{\bf k\cdot }({\bf s}-
{\bf s_0})}G({\bf k},\xi )  \label{freegenf} 
\end{equation} 
and  
\begin{equation} 
P_n^F({\bf s}|{\bf s_0})=\int_{{\bf k}}
\oint_\xi \xi^{-n}e^{-i{\bf k\cdot}
({\bf s}-{\bf s_0})}G({\bf k},\xi )\quad .  \label{freeS} 
\end{equation} 
 
In subsequent sections, several 
probability distributions will occur 
frequently. For convenience, we 
summarize their properties here. Define  
\begin{eqnarray} 
&&t\equiv P^F({\bf 0}|{\bf 0};\xi )\;,  \label{t} \\ 
&&u\equiv P^F({\bf a}|{\bf 0};\xi )\;,  \label{u} \\ 
&&h\equiv P^F({\bf a}|-{\bf a};\xi )\;,  \label{cross} \\ 
&&v\equiv P^F({\bf a}|{\bf b};\xi )
\;,\;\;\;{\bf b}\neq \pm {\bf a} 
\;,\;\;\;(\mbox{only for}\;\;d\geq 2)  \label{simhbt} 
\end{eqnarray} 
all of which are known functions of $\xi$. 
In particular, $t(\xi )$ is just 
the generating function for the return 
probability of a pure random walk. 
The  definitions (\ref{t}), (\ref{cross}) 
and (\ref{simhbt}) allow us to write  
\begin{equation} 
P^F({\bf a}|{\bf b};\xi )=
v+(t-v)\delta _{{\bf a},{\bf b} 
}+(h-v)\delta _{{\bf a},-{\bf b}}\ \quad ,  \label{Pvth} 
\end{equation}
where both ${\bf a}$ and ${\bf b}$ are
 lattice vectors (nearest neighbors
of the origin).
Applying (\ref{sumPbxi}) to the simplest 
random walk ($p=1/2d$), we find 
relations between $t,u,h,$ and $v$ \cite{ZT}, e.g., by choosing
${\bf s}_0 \equiv 0$ in (\ref{sumPbxi}) 
\begin{equation} 
t=1+\xi u\;\;,  \label{uandt} 
\end{equation} 
and (for $d\geq 2$) by choosing 
 ${\bf s}_0 \equiv {\bf b}$ 
\begin{equation} 
\xi p [ t+2(d-1)v+h] =u\;\;.  \label{tvhu} 
\end{equation} 
Finally, since we will be interested in late
 times ($n\rightarrow \infty $) 
corresponding to the limit $\xi \to 1^{-}$,
 we note the following well known 
asymptotic behavior of $t$ \cite{H}:  
\begin{eqnarray} 
t\rightarrow \left\{ 
\begin{array}{c} 
1/\sqrt{2\left( 1-\xi \right) 
}\quad \quad \mbox{ for }d=1 \\  
\frac 1\pi \ln \left( \frac 8{1-\xi }\right) 
\quad \quad \quad \mbox{ for }d=2 
\\  
const.\quad \quad \quad \quad \quad \mbox{ for }d>2 
\end{array} \right. \label{asymp} 
\end{eqnarray} 
From (\ref{uandt}), we conclude that $u$ behaves the same way. 
 
\subsection{Walks with a hop-over site} 
 
Without loss of generality, let us place 
the hop-over site at the origin. 
The probability for the walker to hop
 from ${\bf -a}$ to ${\bf a}$ will be 
denoted by $p_{{\bf a}}$. By keeping 
this rate different from $p$, we will 
be able to log the different hop-over attempts. 
Clearly, $P^H_n({\bf s}|{\bf s_0})$, 
the distribution with such a site, will be a 
polynomial in the various $p_{{\bf a}}$'s. 
Meanwhile, the co-efficient of 
$p_{{\bf a}}^{\nu_{{\bf a}}}$ will 
be associated with the subset of those 
walks which have hopped over the 
origin (from ${\bf -a}$ to ${\bf a}$)
 $\nu_{{\bf a}}$ times. We will return to 
these considerations in more detail below. 

%\begin{figure}[htbp]
%\vspace*{-1.5cm} \hspace*{3cm}
%\epsfxsize = 3.4 in \epsfbox{prop1.eps}  
%\vspace*{-1.2cm}         
%\caption{Random walk with a single hop-over site.
%The diamond symbolizes the hop-over site located
%in the origin and the circle represents the vacancy.}     
%\end{figure}     

Again, we let the walker start at ${\bf s_0}$,
which is not the origin, i.e,  
\begin{equation} 
P_0^H({\bf s}|{\bf s_0})=\delta _{{\bf s},{\bf s_0}}
\quad ,\;\;\;\mbox{and}\;\;\;{\bf  
s_0}\neq {\bf 0}\;\;.  \label{hopinit} 
\end{equation} 
The subsequent evolution of $P^H$ is governed 
by the following Master 
equation:  
\begin{eqnarray} 
P_{n+1}^H({\bf s}|{\bf s_0})-P_n^H({\bf s}|{\bf s_0}) &=&
p(1-\delta _{{\bf s},{\bf 0} 
}) \sum_{\{{\bf a\}}}\left[ 
P_n^H({\bf s+a}|{\bf s_0})-P_n^H({\bf s}|{\bf s_0})\right]  
\nonumber \\ 
&&+\sum_{\{{\bf a}\}}\delta _{{\bf s},{\bf a}}
\left[ p_{{\bf a}}P_n^H(-{\bf a}|{\bf s_0})
-\left(p_{-{\bf a}}-p\right) 
P_n^H({\bf a}|{\bf s_0})\right]  \label{MEhop} 
\end{eqnarray} 
the first term on the right shows that the 
walker never visits the forbidden 
site (${\bf 0}$) and that, away from the
 neighborhood of the origin, it 
performs a simple P\'{o}lya walk (a 
``free walk''). The latter term 
describes the possibility of hop-over, 
when the walker finds itself on a 
nearest neighbor site of ${\bf 0.}$ (See Fig.1.) 
 
Going over to (${\bf k},\xi $) space by 
(\ref{gendefF},\ref{gendefP}), this 
equation becomes  
\begin{eqnarray*} 
G^{-1}({\bf k};\xi )\tilde{P}^H({\bf k};\xi)
-e^{i{\bf k}{\bf s_0}} &=&-\xi 
p\sum_{\{{\bf a\}}}P^H({\bf a}|{\bf s_0};\xi ) \\ 
&&+\xi \sum_{\{{\bf a\}}}\left[ p_{{\bf a}}
P^H(-{\bf a}|{\bf s_0};\xi )-\left( p_{- 
{\bf a}}-p\right) P^H({\bf a}|{\bf s_0};\xi )\right]
 e^{i{\bf k\cdot }{\bf a}}\quad . 
\end{eqnarray*} 
This representation clearly displays 
the effects of the ``defect'' 
(associated with the hop-over site),
 since the left hand side consists of 
the terms for the pure random walk, 
only. A more elegant form would be  
\begin{equation} 
G^{-1}({\bf k};\xi )\tilde{P}^H
({\bf k};\xi )-e^{i{\bf k}{\bf s_0}}=\xi 
\sum_{\{{\bf a\}}}\Gamma 
({\bf k},{\bf a})P^H({\bf a}|{\bf s_0};\xi )  \label{PHkxi} 
\end{equation}

\begin{figure}[htbp]
\vspace*{-1.5cm} \hspace*{3cm}
\epsfxsize = 3.4 in \epsfbox{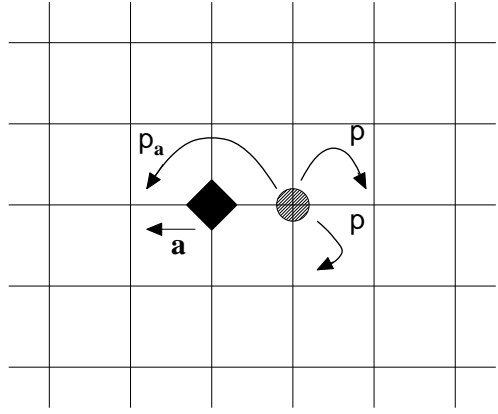}  
\vspace*{-1.2cm}         
\caption{Random walk with a single hop-over site.
The diamond symbolizes the hop-over site located
in the origin and the circle represents the vacancy.}     
\end{figure}     
 
where  
\begin{equation} 
\Gamma ({\bf k},{\bf a})\equiv 
p\left( e^{i{\bf k\cdot }{\bf a}}-1\right) 
+p_{-{\bf a}}\left( e^{-i{\bf k\cdot }
{\bf a}}-e^{i{\bf k\cdot }{\bf a}
}\right)  \label{Gamma} 
\end{equation} 
Now, the right hand side of (\ref{PHkxi}) 
can be regarded as an extra 
inhomogeneity for the solution:  
\begin{equation} 
\tilde{P}^H({\bf k};\xi )=G({\bf k};\xi )
\left[ e^{i{\bf k}{\bf s_0}}+\xi 
\sum_{\{{\bf a\}}}\Gamma ({\bf k},{\bf a})
P^H({\bf a}|{\bf s_0};\xi )\right] \label{PHsol} 
\end{equation} 
 
To find the solution explicitly, we must 
determine the $2d$ quantities $P^H({\bf a}|{\bf s_0};\xi )$.
 This can be done by exploiting 
(\ref{invertS}) and (\ref{PHsol})  
\begin{eqnarray} 
P^H({\bf a}|{\bf s_0};\xi ) &=&\int_{{\bf k}}
e^{-i{\bf k\cdot }{\bf a}}
\tilde{P}^H({\bf k};\xi )  \nonumber  \label{PMP} \\ 
&=&\int_{{\bf k}}e^{-i{\bf k\cdot }
{\bf a}}G({\bf k};\xi )\left[ e^{i{\bf k}
{\bf s_0}}+\xi 
\sum_{\{{\bf b\}}}\Gamma ({\bf k},
{\bf b})P^H({\bf b}|{\bf s_0};\xi 
)\right]   \nonumber \\ 
&=&P^F({\bf a}|{\bf s_0};\xi )+\sum_{\{{\bf b\}}}\xi
\int_{{\bf k}}e^{-i{\bf k\cdot } 
{\bf a}}G({\bf k};\xi )\Gamma 
({\bf k},{\bf b})P^H({\bf b}|{\bf s_0};\xi )  \label{PMP1} 
\end{eqnarray} 
Linear in $P^H({\bf a}|{\bf s_0};\xi )$,
 equation (\ref{PMP1}) can be solved:  
\begin{equation} 
P^H({\bf a}|{\bf s_0};\xi )=
\sum_{\{{\bf b\}}}{\bf L}_{{\bf a},{\bf b}}(\xi )
P^F({\bf b}|{\bf s_0};\xi )  \label{Pa} 
\end{equation} 
where ${\bf L}$ is the inverse of 
the $2d\times 2d$ matrix with elements:  
\begin{equation} 
({\bf L}^{-1})_{{\bf a},{\bf b}}=
\delta _{{\bf a},{\bf b}}-\xi \int_{{\bf k} 
}e^{-i{\bf k\cdot }{\bf a}}G({\bf k};
\xi )\Gamma ({\bf k},{\bf b})\quad . 
\label{Linv} 
\end{equation} 
Once $P^H({\bf a}|{\bf s_0};\xi )$ are known, 
they can be substituted back into (\ref{PHsol}) 
and the explicit solution can be obtained:  
\begin{equation} 
\tilde{P}^H({\bf k};\xi )=
\tilde{P}^F({\bf k};\xi )+\sum_{\{{\bf a},{\bf b} 
\}}\xi G({\bf k};\xi )\Gamma ({\bf k},
{\bf a}){\bf L}_{{\bf a},{\bf b}}(\xi 
)P^F({\bf b}|{\bf s_0};\xi )\quad .  \label{PH} 
\end{equation} 
Using (\ref{invertS}), we may return to the lattice:  
\begin{equation} 
P^H({\bf s}|{\bf s_0};\xi )=P^F({\bf s}|{\bf s_0};\xi )+
\sum_{\{{\bf a},{\bf b}\}}\left[ \xi 
\int_{{\bf k}}e^{-i{\bf k\cdot }{\bf s}}
G({\bf k};\xi )\Gamma ({\bf k},{\bf a 
})\right] {\bf L}_{{\bf a},{\bf b}}(\xi )
P^F({\bf b}|{\bf s_0};\xi )\quad . 
\label{PHI} 
\end{equation} 
Making use of Eq. (\ref{Gamma}) the 
elements of ${\bf L}^{-1}$ become 
expressed solely in free walk terms:  
\begin{equation} 
({\bf L}^{-1})_{{\bf a},{\bf b}}=
\delta _{{\bf a},{\bf b}}-\xi \left\{ p_{- 
{\bf b}}P^F({\bf a}|-{\bf b};\xi )+
(p-p_{-{\bf b}})P^F({\bf a}|{\bf b};\xi )-
pP^F({\bf a}|{\bf 0};\xi )\right\} \quad . 
\label{explGamma} 
\end{equation} 
The integral in the square bracket of
 Eq. (\ref{PHI}) is computed similarly:  
\begin{equation} 
\xi \int_{{\bf k}}e^{-i{\bf k}{\bf s}}
G({\bf k};\xi )\Gamma ({\bf k},{\bf a} 
)=\xi p_{-{\bf a}}P^F({\bf s}|-{\bf a};\xi )+\xi 
(p-p_{-{\bf a}})P^F( 
{\bf s}|{\bf a};\xi )-\xi pP^F({\bf s}|{\bf 0};\xi ) 
\label{hopgen} 
\end{equation} 
Equations (\ref{PHI}-\ref{hopgen}) 
express the generating function of the 
hop-over walk in terms of the hop-over
 rates $\{p_{{\bf a}}\}$ and the 
generating function characterizing 
the free walk. It may be worthwhile to 
interpret the second term in (\ref{PHI}) 
physically. From  (\ref{Linv}) $ 
{\bf L}$ can be expanded as a power
 series in ${\bf \Gamma }$, which 
accounts for the difference between a 
free passage through the origin and an 
hop-over. Thus, the $n^{th}$ term
 in this ${\bf \Gamma }$ series in (\ref 
{PHI}) can be thought of as paths 
which have $n$ encounters with the 
hop-over site. This connection 
will be explored further in the next 
subsection. 
 
Here, we end with a closer look into the
 case with {\em isotropic} hop-over 
rates, i.e.,  
\[ 
p_{{\bf a}}=p^{\prime }\quad , 
\] 
but with $p^{\prime }$ not necessarily 
being equal to $p$. Then, eqns.
 (\ref{PHI}-\ref{hopgen}) reduce to 
\begin{eqnarray} 
P^H({\bf s}|{\bf s_0};\xi )=P^F({\bf s}|{\bf s_0};\xi )+\xi
 \sum_{\{{\bf a},{\bf b}\}} \Big[\;
p^{\prime }P^F({\bf s}|-{\bf a};\xi)+
(p-p^{\prime })P^F({\bf s}|{\bf a};\xi )&&\nonumber \\
- pP^F({\bf s}|{\bf 0};\xi )\Big] {\bf L}_{ 
{\bf a},{\bf b}}(\xi )P^F({\bf b}|
{\bf s_0};\xi )&&.  \label{PHISO} 
\end{eqnarray} 
while our task is to invert,   
\begin{equation} 
({\bf L}^{-1})_{{\bf a},{\bf b}}=
\delta _{{\bf a},{\bf b}}-\xi \left[ 
p^{\prime }P^F({\bf a}|-{\bf b};\xi )
+(p-p^{\prime })P^F({\bf a} 
|{\bf b};\xi )-pP^F({\bf a}
|{\bf 0};\xi )\right] \quad , 
\label{LinvISO} 
\end{equation} 
where ${\bf a}$ and ${\bf b}$ are nearest
 neighbors of the origin. Due to 
isotropy and homogeneity, this matrix 
reduces considerably. The last term in 
the [...] brackets above is seen 
to be a constant (\ref{u}), $-pu$, for  
{\em all} matrix elements. Using 
(\ref{Pvth}), we write (\ref{LinvISO}) as  
\begin{eqnarray} 
({\bf L}^{-1})_{{\bf a},{\bf b}}=\xi
 p(u-v)+\delta _{{\bf a},{\bf b}}\left[ 
1-\xi \left( p^{\prime }(h-t)+p(t-v)
\right) \right] && \nonumber \\
 -\delta _{{\bf a},-{\bf b 
}}\xi \left[ p^{\prime }(t-h)+p(h-v)
\right] &&  \label{Linvp} 
\end{eqnarray} 
 
Since this matrix is of the form  
\begin{equation} 
A+B\delta _{{\bf a},{\bf b}}+
C\delta _{{\bf a},-{\bf b}}\quad ,  \label{ABC} 
\end{equation} 
it is easy to check that its 
inverse is also of this form, namely,  
\begin{eqnarray} 
{\bf L}_{{\bf a},{\bf b}} &=&A^{\prime }
+B^{\prime }\delta _{{\bf a},{\bf b}% 
}+C^{\prime }\delta _{{\bf a},-{\bf b}}  \label{Lp1} \\ 
&=&\frac 1{B^2-C^2}\left[ \left( 
\frac{A(C-B)}{2dA+B+C}\right) +B\delta _{ 
{\bf a},{\bf b}}-C\delta _{{\bf a},
-{\bf b}}\right] \quad .  \label{Lp} 
\end{eqnarray} 
With the help of (\ref{Linvp}-\ref{Lp}),
 we can carry out the matrix 
multiplication in (\ref{PHISO}).
 Note that a quite a few sums over ${\bf  
\{a,b}\}$ decouple. For example, using 
(\ref{sumPbxi}) and $P^F({\bf s} 
|{\bf s}^{\prime };\xi )=
P^F({\bf s}^{\prime }|{\bf s};\xi )$ 
, we have  
\[ 
\xi \sum_{\{{\bf a}\}}
\left[ p^{\prime }P^F({\bf s}|-{\bf a};\xi 
)+(p-p^{\prime })
P^F({\bf s}|{\bf a};\xi )-pP^F({\bf s}|{\bf  
0};\xi )\right] =
\left( 1-\xi \right) P^F({\bf s}|{\bf 0};\xi )\quad 
. 
\] 
The explicit expression for our 
walker to move from $\!{\bf s}_{{\bf 0}}$ to  
${\bf s}$, with isotropic 
hop-over rate $p^{\prime }$, is 
\begin{eqnarray} 
P^H({\bf s}|{\bf s_0};\xi ) 
&=&P^F({\bf s}|{\bf s_0};\xi )
-\left[ B^{\prime }+C^{\prime 
}-A^{\prime }\left( 1-\xi \right) /
\xi p\right] \left[ 1-\xi +2d\xi p\right] 
P^F({\bf s}|{\bf 0};\xi )P^F({\bf 0}|{\bf s}_{{\bf 0}};\xi )  
\nonumber \\ 
&&+\xi \left[ B^{\prime }p+\left(
 C^{\prime }-B^{\prime }\right) p^{\prime 
}\right] \sum_{\{{\bf a}\}}P^F({\bf s}|
{\bf a};\xi )P^F( 
{\bf a}|{\bf s_0};\xi )  \nonumber \\ 
&&+\xi \left[ C^{\prime }p+(B^{\prime }-
C^{\prime })p^{\prime }\right] 
\sum_{\{{\bf a}\}}P^F({\bf s}|-{\bf a};\xi )P^F({\bf a}
|{\bf s_0};\xi ) 
\end{eqnarray} 
where $A^{\prime },$ $B^{\prime}$,
 and $C^{\prime }$ are to be read off 
from (\ref{Linvp}-\ref{Lp}). 
Except for the first term 
(which stands for the `free' case), this
expression shows the non-trivial 
effect of a single hop-over site on the
random walk explicitly .
The various terms can be easily interpreted:
 the second term removes the
(free) walks which pass through the origin, 
while the last two takes into
account walks which land on a neighbor.
If we take
 the usual continuum limit 
(the lattice spacing $a \to 0$, 
unit time-step $\tau \to 0$, such that
$a^2/\tau=2d$) of this expression, 
the effect becomes vanishingly small.
This result is perhaps not surprising, 
since a single site cannot affect
the properties of the random walker 
in the large distance, long time limit.
However, buried in this approach 
is a non-trivial question, namely, the
statistics associated with the 
hop-overs (how often and in which direction
does the walker hops over the 
origin). That is the subject of the next
section.

\subsection{Statistics of hop-overs} 
 
Next, we turn to a study of the 
statistics of the hop-overs. Note that, due 
to the longer jumps (across the origin),
 the walker suffers an {\em extra} 
displacement when compared to the 
simple free walk. We define the variable $% 
\mbox{\boldmath$\rho$}$ as the 
{\em negative} of this extra displacement, 
for reasons that will become obvious 
later. We will also be interested in 
the number of times a hop-over jump
 has occurred. Thus, we define $\phi_n^\nu ({\bf s},
\mbox{\boldmath$\rho$}|{\bf s_0})$ as the 
probability that, starting from ${\bf s_0}$,
 the walker arrives at site 
 ${\bf s}$ after $n$ steps, 
 
\begin{itemize} 
\item  having performed $\nu $ hops over the origin and 
 
\item  suffering a total ``extra'' displacement equal to $- 
\mbox{\boldmath$\rho$}$. 
\end{itemize} 
 
As usual, it is more convenient to work 
with generating functions in Fourier 
space. Accordingly, we trade the variables 
(${\bf s},\mbox{\boldmath$\rho$}$ 
) and ($n,\nu $) for their conjugates: 
(${\bf k},\mbox{\boldmath$\kappa$}$) 
and ($\xi ,\zeta $). We can write:
%Suppressing 
%the dependence on ${\bf s_0}$, we write  
\begin{equation} 
\tilde{\Phi}({\bf k},
\mbox{\boldmath$\kappa$};\xi ,\zeta )=\sum_{n,\nu 
=0}^\infty \sum_{{\bf s},\mbox{{\scriptsize  
 \boldmath$\rho$}}}\xi^n
 \zeta^\nu e^{i({\bf k\cdot }{\bf s}+ 
\mbox{{\scriptsize 
\boldmath$\kappa$}}\cdot 
\mbox{{\scriptsize\boldmath$\rho$}})}
\phi _n^\nu ( 
{\bf s},\mbox{\boldmath$\rho$}|{\bf s_0})\quad .  \label{togen} 
\end{equation} 
Once $\tilde{\Phi}({\bf k},
\mbox{\boldmath$\kappa$};\xi ,\zeta )$ is known, 
the inverse transforms will lead us to the distribution itself:  
\begin{equation} 
\phi _n^\nu({\bf s},\mbox{\boldmath$\rho$}|{\bf s_0})
=\oint_{\xi ,\zeta }\int_{{\bf k} 
,\mbox{{\scriptsize \boldmath$\kappa$}}}
\xi ^{-n}\zeta ^{-\nu }e^{-i({\bf % 
k\cdot }{\bf s}+\mbox{{\scriptsize 
\boldmath$\kappa$}}\cdot  
\mbox{{\scriptsize\boldmath$\rho$}})}
\tilde{\Phi}({\bf k}, 
\mbox{\boldmath$\kappa$};
\xi ,\zeta )\quad .  \label{toback} 
\end{equation} 
 
To arrive at an expression for 
$\tilde{\Phi}({\bf k},\mbox{\boldmath$\kappa$}
;\xi ,\zeta )$, we first study a more
detailed distribution: $\phi _n^{\{\nu 
_{{\bf a}}\}}({\bf s}|{\bf s_0})$, i.e., 
the probability of the walker makes exactly 
$\nu _{{\bf a}}$ hops from $-{\bf a}$ 
to ${\bf a}$ (for each ${\bf a}$). Here  
$\{\nu _{{\bf a}}\}$ denotes the set 
of $2d$ numbers associated with the 
different directions of hop-over. 
Since each hop-over may occur only with 
probability $p_{{\bf a}}$, and the 
events are independent, we conclude that $ 
\phi _n^{\{\nu _{{\bf a}}\}}({\bf s}|{\bf s_0})$
 must contain the factor $\prod_{\{ 
{\bf a\}}}p_{{\bf a}}^{\nu _{{\bf a}}}$,
 where $\Pi _{\{{\bf a\}}}$ is 
product over all the $2d$ nearest neighbors. 
 
Meanwhile, the extra displacement is just 
$-\mbox{\boldmath$\rho$}=\sum_{\{ 
{\bf a\}}}{\bf a}\nu _{{\bf a}}$, so that  
\begin{equation} 
\phi_n^\nu ({\bf s},\mbox{\boldmath$\rho$}|{\bf s_0})
=\sum_{\{\nu _{{\bf a}}{\bf \}} 
}\phi _n^{\{\nu _{{\bf a}}\}}({\bf s}|{\bf s_0})\;
\delta _{\nu ,\sum_{\{{\bf a\}}}\nu 
_{{\bf a}}}\;\delta _{-\mbox{{\scriptsize 
\boldmath$\rho$}},\sum_{\{{\bf a\}} 
}{\bf a}\nu _{{\bf a}}}\quad .  \label{relatem} 
\end{equation} 
In terms of generating functions, this expression becomes:  
\begin{equation} 
\tilde{\Phi}({\bf k},\mbox{\boldmath$\kappa$};
\xi ,\zeta )=\sum_{\{\nu _{ 
{\bf a}}{\bf \}}}\tilde{\Phi}^{\{
\nu _{{\bf a}}\}}({\bf k};\xi )\prod_{\{ 
{\bf a\}}}\zeta ^{\nu _{{\bf a}}}
e^{-i\mbox{{\scriptsize \boldmath$\kappa$}} 
\cdot {\bf a}\nu _{{\bf a}}}  \label{karacsony} 
\end{equation} 
where $\tilde{\Phi}^{\{\nu _{{\bf a}}
\}}({\bf k};\xi )$ is the generating 
function and Fourier transform of 
$\phi _n^{\{\nu _{{\bf a}}\}}({\bf s}|{\bf s_0})$. 
 
Armed with these considerations, we 
interpret the expression (\ref{PH}) 
physically. Expanding ${\bf L}$ in a 
power series in the $p_{{\bf a}}$'s, we 
see that $P^H$ itself is a power series
 in these rates. Naively, it is 
tempting to identify these co-efficients 
with $\phi _n^{\{\nu _{{\bf a}}\}}(% 
{\bf s})$. However, there is an {\em 
implicit} probability for the walker to  
{\em remain} at the site ${\bf -a}$ :  
\[ 
\left( 1-p_{{\bf a}}-(2d-1)p\right) \quad , 
\] 
so that the co-efficient of $p_{{\bf a}}$ 
also includes paths that avoid the 
origin. Another way to see this 
difficulty is through the original Master 
equation (\ref{MEhop}), in which a 
particular $p_{{\bf a}}$ appears twice, 
once in the ``gain'' of $P^H({\bf a})$ 
and once in the ``loss'' of $P^H(- 
{\bf a})$. Of course, both of these 
terms represent the {\em same} jump. 
Therefore, to avoid double counting, 
we can label one of them by $q_{{\bf a}} 
$ and, only at the end, set 
$q_{{\bf a}}=p_{{\bf a}}$. Let us emphasize 
that, with $q_{{\bf a}}\neq p_{{\bf a}}$,
 total probability will not be 
conserved. But this is precisely the 
trick for distinguishing making a 
hop-over jump from {\em not} making one. 
 
For definiteness, let us choose to 
label the ``gain'' term by $q_{{\bf a}}$. 
A little care leads us to a 
modified version of (\ref{Gamma}):  
\begin{eqnarray} 
\overline{\Gamma }({\bf k},{\bf a}) 
&\equiv &p\left( e^{i{\bf k\cdot }{\bf a} 
}-1\right) +q_{-{\bf a}}e^{-i{\bf k
\cdot }{\bf a}}-p_{-{\bf a}}e^{i{\bf % 
k\cdot }{\bf a}}  \nonumber  \label{Gammatilde} \\ 
&=&q_{-{\bf a}}e^{-i{\bf k\cdot }
{\bf a}}+\left( p-p_{-{\bf a}}\right) e^{i 
{\bf k\cdot }{\bf a}}-p  \label{Gammatilde2} 
\end{eqnarray} 
from which we can find  
\begin{equation} 
\tilde{\overline{P}}{\;\!}^{H}({\bf k};\xi )
=\tilde{P}^F({\bf k};\xi )+\sum_{\{ 
{\bf a},{\bf b}\}}\xi G({\bf k};\xi)
\overline{\Gamma }({\bf k},{\bf a}) 
\overline{L}_{{\bf a},{\bf b}}(\xi )
P^F({\bf b};\xi )\quad .  \label{PHtilde} 
\end{equation} 
Note that the overline over a 
quantity symbolizes the dependence on both 
sets of rates: $\{q_{{\bf a}}\}$ and $\{p_{{\bf a}}\}$. 
 
With this trick, we can associate 
the co-efficient of $\left( q_{{\bf a} 
}\right) ^{\nu _{{\bf a}}}$ , in 
a power series expansion of 
$\tilde{\overline{P}}{\;\!}^{H}$, with walks 
that include $\nu _{{\bf a}}$ hop-overs 
(jumps from $-{\bf a}$ to ${\bf a}$). 
Now, we must pick out these 
co-efficients and then multiply them
with the correct weight: $\left( p_{ 
{\bf a}}\right)^{\nu _{{\bf a}}}$. 
This is accomplished by applying $p_{ 
{\bf a}}^{\nu _{{\bf a}}}
\oint_{q_{{\bf a}}}q_{{\bf a}}^{-\nu _{{\bf a}}}$ 
for each ${\bf a}$, so that  
\begin{equation} 
\tilde{\Phi}^{\{\nu _{{\bf a}}\}}
({\bf k};\xi )=\prod_{\{{\bf a\}}}\oint_{q_{ 
{\bf a}}}\left( p_{{\bf a}}/q_{{\bf a}}
\right)^{\nu _{{\bf a}}}
\tilde{\overline{P}}{\;\!}^{H}
({\bf k};\xi )\quad .  \label{invertNa} 
\end{equation} 
 
Substituting Eq. (\ref{invertNa}) back 
into (\ref{karacsony}), we see that 
the dependence on the variables $\nu _{{\bf a}}$ 
factorizes. Performing the 
sums over each $\nu _{{\bf a}}$, we arrive at:  
\begin{equation} 
\tilde{\Phi}({\bf k},\mbox{\boldmath$\kappa$};
\xi ,\zeta )=\prod_{\{{\bf a\}} 
}\frac 1{2\pi i}\oint \frac{dq_{{\bf a}}}
{q_{{\bf a}}-p_{{\bf a}}\zeta e^{-i 
\mbox{{\scriptsize \boldmath$\kappa$}}\cdot 
{\bf a}}}\tilde{\overline{P}}{\;\!}^{H}
({\bf k};\xi )\quad .  \label{calcul} 
\end{equation} 
i.e.,  
\begin{equation} 
\tilde{\Phi}({\bf k},\mbox{\boldmath$\kappa$};
\xi ,\zeta )=\tilde{\overline{P}}{\;\!}^{H}({\bf k};\xi )
\bigg|_{q_{{\bf a}}=p_{{\bf a}}\zeta e^{-i 
\mbox{{\scriptsize 
 \boldmath$\kappa$}}\cdot {\bf a}}}\quad .  \label{replace} 
\end{equation} 
In other words, $\tilde{\Phi}({\bf k},
\mbox{\boldmath$\kappa$};\xi,\zeta )$ 
is nothing but $\tilde{\overline{P}}{\;\!}^{H}
({\bf k};\xi )$ with all the $q_{ 
{\bf a}}$'s simply replaced by 
$p_{{\bf a}}\zeta e^{-i\mbox{{\boldmath$ 
\kappa$}}{\bf a}}$. 
 
From expression (\ref{PHtilde}) for
 $\tilde{\overline{P}}{\;\!}^{H}$, we write 
explicitly:  
\begin{equation} 
\tilde{\Phi}({\bf k},\mbox{\boldmath$\kappa$};
\xi ,\zeta )=\tilde{P}^F({\bf k 
};\xi )+\sum_{\{{\bf a},{\bf b}\}}\xi 
G({\bf k};\xi )\overline{\Gamma }({\bf  
k},\mbox{\boldmath$\kappa$},
{\bf a};\zeta )\overline{{\bf L}}_{{\bf a},{\bf b 
}}(\mbox{\boldmath$\kappa$};\xi
 ,\zeta )P^F({\bf b}|{\bf s_0};\xi )\;,  \label{elso} 
\end{equation} 
where  
\begin{equation} 
\overline{\Gamma }({\bf k},\mbox{\boldmath$\kappa$},
{\bf a};\zeta )=p_{-{\bf  
a}}\left[ \zeta e^{-i({\bf k}-\mbox{{
\scriptsize \boldmath$\kappa$}})\cdot  
{\bf a}}-e^{i{\bf k\cdot }{\bf a}}
\right] +p\left( e^{i{\bf k\cdot }{\bf a} 
}-1\right)  \label{repGt} 
\end{equation} 
and  
\begin{equation} 
(\overline{{\bf L}}^{-1}
(\mbox{\boldmath$\kappa$};\xi ,\zeta ))_{{\bf a}, 
{\bf b}}=\delta _{{\bf a},{\bf b}}
-\xi \left\{ p_{-{\bf b}}\zeta e^{i 
\mbox{{\scriptsize \boldmath$\kappa$}}
\cdot {\bf b}}P^F({\bf a}|- 
{\bf b};\xi )+(p-p_{-{\bf b}})
P^F({\bf a}|{\bf b};\xi )-pu\right\} 
\label{explGammaP} 
\end{equation} 
 
Due to the extra factor $\zeta 
e^{i\mbox{{\scriptsize \boldmath$\kappa$}} 
\cdot {\bf b}}$, it is difficult, 
even for the isotropic hop-over walk, to 
invert $\overline{{\bf L}}$ in 
arbitrary $d.$ However, since most of the 
interesting effects occur in $d=2$ 
(where the return probability approaches 
unity for large times), the 
$4\times 4$ matrix can always be inverted. 
 
\subsection{A hidden pure random walk} 
 
Finally, if we are interested in the simple
hop-over problem in which all 
the rates are identical, then we simply 
set all $p_{{\bf a}}$'s to $p$. In 
this case, we should also retrieve the
pure random walk if we {\em subtract }
the extra displacements due to the hop-overs.
 In other words, let us 
consider the probability of finding the 
walker at ${\bf s-}\left(-\mbox{{\scriptsize 
\boldmath$\rho$}}\right)$, regardless of 
how many hop-overs occured or 
their directions. Thus, we study    
\[ 
P_n({\bf r}|{\bf s_0})\equiv \sum_{{\bf s}}
\sum_{\mbox{{\scriptsize \boldmath$\rho$}} 
}\delta _{{\bf r},{\bf s}+\mbox{{\scriptsize 
\boldmath$\rho$}}}\;\sum_{\nu =0}^\infty \phi_n^\nu ({\bf s}, 
\mbox{\boldmath$\rho$}|{\bf s_0})\quad . 
\] 
Again, it is easier to study the generating
 function in Fourier space: Using  
${\bf k}$ as the variable conjugate to 
${\bf r}$, we see that the above sums 
result in:  
\begin{eqnarray*} 
\tilde{P}({\bf k};\xi ) &=&\sum_{n,{\bf r}}
e^{i{\bf k\cdot r}}\xi ^nP_n({\bf 
r}|{\bf s_0}) \\ 
&=&\tilde{\Phi}({\bf k},\mbox{\boldmath$\kappa$}
={\bf k};\xi ,\zeta =1)\quad . 
\end{eqnarray*} 
Setting in (\ref{repGt}) $p_{-{\bf a}}=p$
 for each ${\bf a}$, $\mbox{\boldmath$\kappa$}=
 {\bf k}$ and $\zeta =1$ we see that $\overline{ 
\Gamma }({\bf k},\mbox{\boldmath$\kappa$}
={\bf k},{\bf a};\zeta =1)=0$. 
Thus, from (\ref{PHtilde}), we obtain  
\begin{equation} 
\tilde{P}({\bf k};\xi )=
\tilde{P}^F({\bf k};\xi )\quad ,  \label{PIPF} 
\end{equation} 
justifying our expectation that, if we 
{\em subtract} the displacements due 
to hop-overs, the simple random walk re-emerges. 
 
\section{Relation to Tagged Diffusion and Some Applications} 
 
The problem of tagged diffusion has 
been solved previously \cite{BH} (in two-dimensions),\cite{ZT} 
(in $d$-dimensions), 
using a ``direct'' approach, i.e., by 
keeping track of all the possible ways 
for the tag to be displaced by a 
random walker. In this section, we will 
first show that the above analysis can 
be applied to provide a new approach 
to this venerable problem. As examples 
of applications of our results, we 
devote the latter subsections to the 
dynamics of two particular physical 
systems. 
 
\subsection{The passive walk or tagged diffusion} 
 
For completeness, let us describe tagged 
diffusion in terms of a ``passive'' 
random walker. Returning to our infinite 
$d$-dimensional hypercubic lattice, 
we place two walkers, one active and one 
passive. Respectively, we label 
them B (for Brownian) and T (for tagged, 
or tracer, particle). The former 
performs the simplest random walk, as in 
Section 1.1 with $p=1/2d$. When B 
attempts to move to the site occupied by 
T, the two simply exchange places. 
Thus, T does not move except when it is 
``kicked'' by B. In this sense, the 
motion of T could be called ``passive'' 
and will be referred to as a ``{\em  
Brownian driven walk}''. The mathematical
 properties of this pair of walkers 
are contained in the joint probability  
\begin{equation} 
\Phi _n^\nu ({\bf r},\mbox{\boldmath$\rho$}
|{\bf r_0,0})  \label{Phi(r,rho)} 
\end{equation} 
for finding B at site ${\bf r}$ and T at 
site $\mbox{\boldmath$\rho$}$, on 
the $n$-th step of B and the $\nu $-th 
step of T. Note that, $\nu $ just 
represents the number of ``kicks'' T 
received from B. The last arguments 
refer to the initial condition, i.e., 
T being at the origin ${\bf 0}$ and B 
at site ${\bf r_0}\neq {\bf 0}$. 
 
From the description above, it is clear 
that, {\em relative to} T, B is 
performing a random walk with a hop-over
 site. Thus,  
\begin{equation} 
\Phi _n^\nu ({\bf r},\mbox{\boldmath$\rho$}|{\bf r}_0,
{\bf 0})=\phi _n^\nu ( 
{\bf s},\mbox{\boldmath$\rho$}|{\bf s}_0)  \label{Phi=phi} 
\end{equation} 
of the previous section, provided 
${\bf r_0}\equiv{\bf s_0}$, and  
\begin{equation} 
{\bf s=r-}\mbox{\boldmath$\rho$}  \label{srrho} 
\end{equation} 
which is just the position of B relative 
to T. Of course, we expect the 
results following this approach to be 
identical to those from the more 
``direct'' approaches in \cite{ZT} and 
\cite{BH}. To be brief, here we will 
present only a certain projection of 
the distribution $\Phi _n^\nu ({\bf r}, 
\mbox{\boldmath$\rho$}|{\bf r}_0,{\bf 0})$,
 and compare the results with 
those in Ref. \cite{ZT}. 
 
Let us focus on the probability that 
T received $\nu $ `kicks' during $n$ 
-steps of B, regardless of the final 
locations of the two walkers. Denoting 
this quantity by $\phi _n^\nu ({\bf r}_0)$, it is:  
\begin{equation} 
\phi _n^\nu ({\bf r}_0)=\sum_{{\bf r,}
\mbox{{\scriptsize \boldmath$\rho$}} 
}\Phi _n^\nu ({\bf r},\mbox{\boldmath$\rho$}|
{\bf r}_0,{\bf 0})=\sum_{{\bf s,
}\mbox{{\scriptsize 
\boldmath$\rho$}}}\phi _n^\nu ({\bf s},
\mbox{\boldmath$\rho$}|{\bf s}_0)\ . 
\label{38} 
\end{equation} 
%As before, we will suppress the dependence
%on initial position unless its 
%role needs emphasis. 
 
From (\ref{togen}) and (\ref{replace}),
 we see that the generating function  
\[ 
\phi (\xi ,\zeta )\equiv \sum_{n,\nu }
\phi _n^\nu \xi ^n\zeta ^\nu  
\] 
is given by  
\begin{equation} 
\phi (\xi ,\zeta )=\tilde{\Phi}({\bf 0},
{\bf 0};\xi ,\zeta )=\tilde{ 
\overline{P}}{\;\!}^H({\bf 0},
{\bf 0};\xi ,\zeta )\ .  \label{hely} 
\end{equation} 
 
Since B performs a random walk, we set
 $p_{{\bf a}}=p$ everywhere. Thus, 
from eqn. (\ref{elso}), we have:  
\begin{equation} 
\phi (\xi ,\zeta )=\frac 1{1-\xi }
\left[ 1+\xi \sum_{\{{\bf a},{\bf b}\}} 
\overline{\Gamma }({\bf 0},{\bf 0},{\bf a};
\zeta )\overline{L}_{{\bf a},{\bf  
b}}({\bf 0};\xi ,\zeta )P^F({\bf b}|{\bf s_0};\xi )\;.\right]  
\end{equation} 
Using (\ref{repGt}), (\ref{explGammaP})
 and (\ref{Pvth}), we find:  
\begin{equation} 
\overline{\Gamma }({\bf 0},{\bf 0},{\bf a}
;\zeta )=-p(1-\zeta )  \label{siG} 
\end{equation} 
and  
\begin{equation} 
(\overline{{\bf L}}^{-1}({\bf 0};
\xi ,\zeta ))_{{\bf a},{\bf b}}=\xi p\left( 
u-\zeta v\right) +\left[ 1-\xi \zeta 
p(h-v)\right] \delta _{{\bf a},{\bf b} 
}-\xi \zeta p(t-v)\delta _{{\bf a},-{\bf b}}\;\;.  \label{simpLe} 
\end{equation} 
Following (\ref{ABC}-\ref{Lp}), we write
 ${\bf \bar{L}}_{{\bf a},{\bf b}}$ 
in the form  
\begin{equation} 
{\bf \bar{L}}_{{\bf a},{\bf b}}=A^{\prime }
+B^{\prime }\delta _{{\bf a},{\bf % 
b}}+C^{\prime }\delta _{{\bf a},-{\bf b}}\quad ,  \label{ABC'} 
\end{equation} 
so that  
\begin{equation} 
\phi (\xi ,\zeta )=\frac 1{1-\xi }\left[ 
1-\xi p(1-\zeta )\left( 2dA^{\prime 
}+B^{\prime }+C^{\prime }\right) 
\sum_{\{{\bf b}\}}P^F({\bf b}|{\bf s_0};\xi 
)\;.\right]  
\end{equation} 
Now, to simplify this some more, we make 
use of (\ref{sumPbxi}) and carry 
out the algebra for $A^{\prime },A,$ etc.:  
\begin{equation} 
\frac 1{1-\xi }\left[ 1-\frac{(1-\zeta )}
{1+\xi u+\zeta \xi p\left( 
(2-2d)v-h-t)\right) }P^F({\bf 0}|{\bf s_0};\xi )\right] \;. 
\end{equation} 
Finally, exploiting (\ref{uandt},\ref{tvhu}),
 we write a compact form for 
the generating function: 
 
\begin{equation} 
\phi (\xi ,\zeta )=\frac 1{1-\xi }
\left[ 1-\frac{\xi \left( 1-\zeta \right)  
}{\xi t-\zeta t+\zeta }P^F({\bf 0}|
{\bf s_0};\xi )\right] \quad.\label{verynice} 
\end{equation} 
 
Before inverting this result to 
obtain $\phi _n^\nu $, let us explore the 
scaling properties of $\phi (\xi ,
\zeta )$, in the limit $\xi ,\zeta 
\rightarrow 1^{-}$. For $d\leq 2$, 
$t$ diverges according to (\ref{asymp}), 
but $\left( 1-\xi \right) t\rightarrow 
0$. So does $P^F({\bf 0}|{\bf  
s_0};\xi )$ for any ${\bf s_0}$, 
with the result that $P^F({\bf 0}| 
{\bf s_0};\xi )/t\rightarrow 1$. 
Keeping the leading non-trivial orders, a 
simple scaling function emerges 
 
\begin{equation} 
\left( 1-\xi \right) \phi (\xi ,
\zeta )\rightarrow \frac 1{1+x}, 
\label{scalingphi} 
\end{equation} 
where  
\begin{equation} 
x\equiv \left( 1-\zeta \right) t  \label{scalingvar} 
\end{equation} 
is the scaling variable. If we trace 
the origins of (\ref{asymp}), we can 
use an explicit and more general form:
 $x=$ $\left( 1-\zeta \right) \left( 
1-\xi \right) ^{\left( d-2\right) /2}$. 
 
Returning to the inverse transforms, 
the one with respect to $\zeta $ is 
trivial. Keeping in mind (\ref{uandt}),
 we find that $\oint_\zeta \zeta 
^{-\nu }\phi (\xi ,\zeta )$ is identical
 to eqns. (34) and (35) of Ref. \cite 
{ZT} obtained via a completely different 
method. Since $t$ is a non-trivial 
function of $\xi $, the transform with 
respect to $\xi $ cannot be carried 
out explicitly. Thus, we leave the final
 result in the form of an inverse 
transform: 
 
\noindent    
$\nu \geq 1:$  
\begin{equation} 
\phi _n^\nu ({\bf r}_{{\bf 0}})=\oint_\xi
 \frac{\xi ^{-n-\nu }}{\left( 1-\xi 
\right) t}\left( 1+\xi t-t\right) \left(
 1-1/t\right) ^{\nu -1}P^F({\bf 0} 
|{\bf r_0};\xi )\;,  \label{vegnn} 
\end{equation} 
$\nu =0:$  
\begin{equation} 
\phi _n^0({\bf r}_{{\bf 0}})=\oint_\xi
 \frac{\xi ^{-n}}{\left( 1-\xi \right) 
t}\left[ t-P^F({\bf 0}|{\bf r}_{{\bf 0}};\xi )\right] . 
\label{vegzn} 
\end{equation} 
These are the explicit formulae, in any $d$,
 for the probability that the 
tag particle (passive walker) moves $\nu $
 steps while the vacancy (active 
walker) takes $n$ steps, starting at 
$\!{\bf r}_{{\bf 0}}$ from the tag. Of 
course, for $\nu ,n\rightarrow \infty $, 
the scaling analysis above implies 
that  
\[ 
\nu \sim \sqrt{n}\mbox{ \quad and\quad }\sqrt{\ln n}  
\] 
in $d=1$ and $2$, respectively. Above 
two dimensions, the return probability 
of the random walker remains less than 
unity as $n\rightarrow \infty $, so 
that $\phi _n^\nu $ decays exponentially in $\nu $. 
 
\subsection{Vacancy mediated disordering 
of an A-B alloy with extreme 
anisotropy} 
 
In some binary alloys, an ordered state 
consists of A and B atoms occupying 
alternate sites on a cubic lattice. 
Under appropriate conditions, an atom 
cannot move unless a vacancy comes in 
contact and exchanges places with it
(also coined as the `vacancy-mechanism')
\cite{Sholl}-\cite{Yuri} . 
Therefore, each atom can be regarded 
as a passive walker, while the vacancy 
plays the role of an active walker. 
More precisely, consider a monolayer, 
composed of A and B atoms in a square
 lattice, adsorbed on a substrate which 
interacts strongly with these atoms 
in such a way that the ground state (of 
the monolayer) is a simple checkerboard
 configuration (antiferromagnetic 
state, in the spin language). To 
simplify the problem further, we suppose 
that the {\em intra}layer interactions 
are negligible. A similar arrangement 
may be realised by a binary alloy in 
a NaCl structure on a simple cubic 
lattice, with extreme anisotropic 
interactions. If the interactions are much 
stronger along the c-axis and we are 
focusing on a (0,0,1) surface, the 
surface atoms will experience a much 
larger interaction with the bulk than 
with their neighbors on the surface. 
For clarity, we will distinguish the 
surface atoms from bulk ones, through
 labeling the former by A/B and the 
latter by a/b. The ground state, is 
shown in Fig. 2a, with only Ab and Ba 
bonds (and, of course, AB and ab ones).
 Given our assumptions, Bb or Aa 
bonds, considered as ``mismatches'', 
will be more costly. Thus the movement 
of the vacancy can cause such excitations
 (Fig. 2b). In a typical system, 
the vacancy will not perform a pure 
random walk, since its movements will be 
governed by these excitation (or 
de-excitation) energies. However, if 
subjected to sufficiently high 
temperatures, we could neglect this 
dependence and approximate the 
vacancy by a Brownian particle. 

\begin{figure}[htbp]
\vspace*{-1.5cm} \hspace*{3cm}
\epsfxsize = 3.4 in \epsfbox{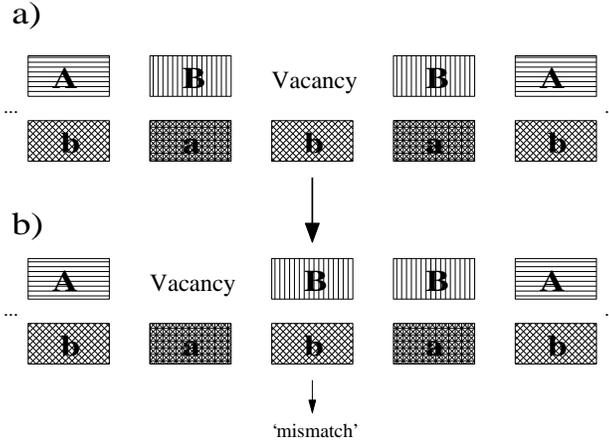}  
\vspace*{-1.3cm}         
\caption{A binary alloy model with extreme anisotropy.
 The configuration a) ``A on top of
b'' or ``B on top of a'' is energetically 
favorable to, e.g., the configuration
 b) ``B on top of b''
 which is considered as a mismatch.}     
\end{figure}
 
Focusing on such a vacancy wandering in 
our monolayer, we can investigate 
the evolution of the excitation energy
 through the creation of the 
mismatches. In particular, the analysis
 above can be applied to answer the 
following question: after a long time 
($n\gg 1$) what is the expected 
excitation energy caused by this vacancy?
 Being proportional to the expected 
number of ``mismatches'', this energy 
shift can be found readily. Note that, 
though the discussion above is clearly 
based on a $d=2$ surface adsorbed on 
a $d=3$ bulk, our considerations 
will be applicable in any $d.$ 
 
First, let the {\em vacancy} be located 
initially at the origin. Then we tag  
{\em each} atom simply by its initial
 location. Alternatively, using (\ref 
{srrho}), we may use ${\bf s_0}$ 
(instead of $-{\bf s_0}$) to label the 
particle {\em uniquely}. After 
the vacancy has taken $n$ steps, atom `${\bf  
s_0}$' has suffered $\nu $ displacements
 with probability $\phi _n^\nu ({\bf  
s_0})$, which is found in (\ref{38}).
 If $\nu $ is odd, there is a 
``mismatch'' for that particle. The 
{\em expected} total number of 
mismatches is therefore expressed by:  
\begin{equation} 
\langle Y_n\rangle =\sum_{{\bf s_0}}\!\!\!{}
\mbox{ }^{\prime }\;\sum_{\nu 
=0}^\infty \frac 12\left[ 1-(-1)^\nu 
\right] \phi _n^\nu ({\bf s_0})\;, 
\label{mismnr} 
\end{equation} 
where the prime on the summation symbol 
means that ${\bf s_0}={\bf 0}$ is 
excluded. Such a sum for the quantities 
in eqns. (\ref{vegnn}) and (\ref 
{vegzn}) is trivial:  
\begin{equation} 
\sum_{{\bf s_0}}\!\!\!\mbox{ }^{\prime }
\;P({\bf 0}|{\bf s_0};\xi 
)=\sum_{{\bf s_0}}P({\bf 0}|{\bf s_0};\xi )-P({\bf 0}|{\bf 0} 
;\xi )=\frac 1{1-\xi }-t\;.  \label{sumr0} 
\end{equation} 
After summing over $\nu $, we find the desired result:  
\begin{equation} 
\langle Y_n\rangle =\oint_\xi \frac{\xi^{1-n}}
{\left( 1-\xi \right) ^2} 
\left[ \frac{\left( \xi -1\right) t+1}
{\left( \xi +1\right) t-1}\right] \;. 
\label{yn} 
\end{equation} 
 
To proceed, let us consider the large 
$n$ limit, corresponding to $\xi \to 
1^{-}$. Since $\left( \xi -1\right) 
t\rightarrow 0$ even for cases when $t$ 
diverges, the $\left[ .\right] $ bracket
 can be replaced by $1/\left( 
2t-1\right) $. The $n\rightarrow \infty $
 behavior can then be extracted by 
exploiting the Discrete Tauberian Theorem
 (See, e.g., Ref. \cite{H}, pp. 118.).

In $d=1$, where we have a linear 
chain of atoms (adsorbed on a two 
dimensional bulk), we find that 
the expected number of mismatches grows 
like:  
\begin{equation} 
\langle Y_n\rangle \sim \sqrt{n}\sqrt{\frac 2\pi }\;. 
\end{equation} 
It is interesting to compare this 
result to the {\em expected number of 
distinct sites} visited in an $n$-step walk,
 which is a well known quantity  
\cite{H}. In $d=1$, it grows like:  
\begin{equation} 
\langle S_n\rangle \sim \sqrt{n}
\sqrt{\frac 8\pi },\;\;\;\mbox{as} 
\;\;\;n\rightarrow \infty \;.  \label{sn1d} 
\end{equation} 
Thus, we find that  
\begin{equation} 
\frac{\langle Y_n\rangle }{\langle
 S_n\rangle }\rightarrow \frac{1}{2} \label{wow1d} 
\end{equation} 
as $n\rightarrow \infty $, i.e., {\em half}
 of the visited sites will have 
mismatches. These results are hardly 
surprising, since, at any time, only 
the string of atoms between the walker 
and its initial position are 
``mismatched''. 
 
For $d=2$, the physical case, the 
results are more interesting. With the 
help of the same theorem, we find:  
\begin{equation} 
\langle Y_n\rangle \sim \frac{\pi}{2}\frac{n}
{\ln {(8n)}},\;\;\;\mbox{as} 
\;\;\;n\rightarrow \infty \;,  \label{y1n2d} 
\end{equation} 
whereas \cite{H}  
\begin{equation} 
\langle S_n\rangle \sim \frac{\pi n}{\ln {(8n)}}\;,\;\;\;\mbox{as} 
\;\;\;n\rightarrow \infty \;,  \label{sn2d} 
\end{equation} 
so that (\ref{wow1d}) proves to be valid in this case also. 
 
Though bulk materials in 4 or higher 
dimensions are not physical, we may 
nevertheless consider the number of 
expected mismatches in $d\geq 3$. Here, $ 
t\left( 1\right) $ is finite, leading to  
\begin{equation} 
\langle Y_n\rangle \sim n\;\frac{1}{2t(1)-1}\;.  \label{yn3d} 
\end{equation} 
Comparing it with  
\begin{equation} 
\langle S_n\rangle \sim n\;
\frac{1}{t(1)}\;,  \label{sn3d} 
\end{equation} 
we find, instead of (\ref{wow1d}), 
a more ``interesting'' result:  
\begin{equation} 
\frac{\langle Y_n\rangle }{\langle 
S_n\rangle }\sim \frac{t(1)}{2t(1)-1} 
=\frac 1{1+R({\bf 0})}\;,\;\;\;
\mbox{as}\;\;n\rightarrow \infty \; \label{wow3d} 
\end{equation} 
where $R({\bf 0})$ is the probability 
that the walker ever returns to its 
starting site, another well known 
quantity in the theory of random walks. 
Actually, eqn. (\ref{wow3d}) is in 
fact valid for any $d\geq 1.$ For $d\leq 
2 $, the walk is recurrent, i.e., 
$R({\bf 0})=1$, reducing (\ref{wow3d}) to 
1/2. Only for $d>2$, is the walks 
transient, where $R({\bf 0})<1$. 
An alternative
way to display (\ref{wow3d}) is to quote the 
ratio of particles performing
an even number of exchanges with the hole 
to those ``visited'' an odd number
of times. Clearly, this ratio is simply $R({\bf 0})$.
These 
remarks give an intuitive explanation
 for (\ref{wow3d}). Since the random 
walk is recurrent in $d=1,2$, every 
particle is sure to be {\em revisited.} 
Therefore, there should be as many 
particles visited an odd number of times 
as those visited an even number of 
times, on the average. However, for $ 
d\geq 3$, a certain amount of the 
particles will be visited only {\em once}. 
Thus, the particles visited an
 odd number of times should be larger. 
 
\subsection{Propagation of interfacial disorder} 
 
As another example of how our 
results can be applied in disordering 
dynamics, we investigate the 
asymptotic behavior of the disorder induced by 
a single Brownian vacancy \cite{TK}. Since 
the most interesting case is $d=2$, we will 
limit our study here. The initial 
configuration is a completely phase 
segregated system, i.e., an infinite
 square lattice, with the upper half 
plane filled with one type of particles 
(white) and the lower half with the 
other type (black). 
A single vacancy (the active random walker) is placed in 
the white region at the interface between
 the two half planes (Fig. 3), and 
labeled as the origin of our 
coordinate system. As the vacancy wanders, 
particles will be drawn into the opposite
 phases, leading to disordering. In 
the following, we will show that, 
after $n$ steps taken by the wanderer, the 
``disorder-profile'' parallel and 
perpendicular to the interface scales as $ 
\sqrt{n}$ and $\sqrt{\ln {n}}$ , 
respectively. The scaling functions are 
combinations of exponentials and a modified Bessel function.

\begin{figure}[htbp]
\vspace*{-0.5cm} \hspace*{3cm}
\epsfxsize = 3.4 in \epsfbox{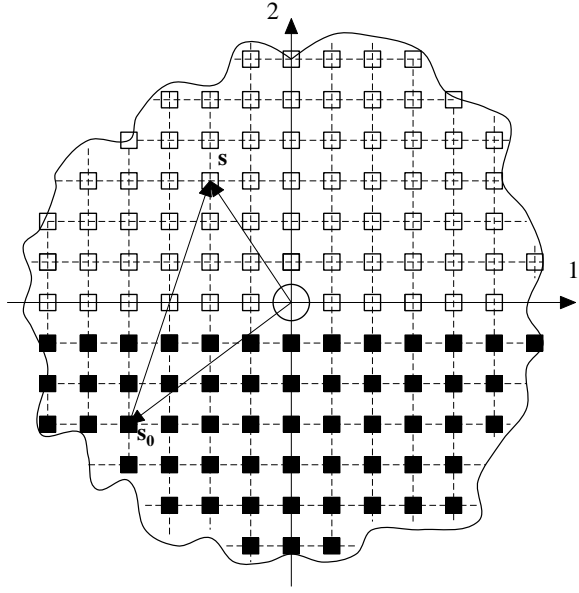}  
\vspace*{-0.5cm}         
\caption{Initial configuration of a sharp interface
between two different species (transparent and filled
squares). The vacancy (large empty circle) is located
initially at the origin of the coordinate system.}     
\end{figure}     
 
Let us focus on the black particles. 
One definition of disorder is the 
density of black particles in the 
upper half plane. Thus, we seek  
\[ 
\Phi_n^{*}({\bf s}) 
\] 
the probability of finding a black
 particle at location ${\bf s}$, 
after $n$ steps of the vacancy. 
Since each particle performs a 
``passive'' random walk independent
 of the others, we may simply track the 
movements of each black particle and 
sum over all those which reach ${\bf s}$. 
As in previous sections each black particle can be 
labeled uniquely by its initial position ${\bf s_0}$.
 
Since we are not interested in neither 
the location of the vacancy nor 
the frequency of a particle being kicked it is clear that
we should sum over the vacancy's position and the frequency
of the hits the black particle receives. 
Summing over all the black particles, we see that  
\begin{equation} 
\Phi _n^{*}({\bf s})=
\sum_{{\bf s_0}}^{(-)}
\phi_n({\bf s}-{\bf s_0}
\!\mid\!-{\bf s_0}) = 
\sum_{{\bf s_0}}^{(-)} 
\sum_{{\bf r}}
\sum_{\nu=0}^{\infty} 
\phi_n^{\nu}({\bf r},{\bf s}-{\bf s_0}
\!\mid\!-{\bf s_0}), \label{sum0}
\end{equation} 
where $\sum^{(-)}$ denotes 
summation over the ${\bf s_0}$'s in
 the lower half plane only
, and ${\bf r}$ is the walker's
position after $n$ steps relative 
to the site ${\bf s_0}$.
In terms of the generating function 
(c.f. eqns. (\ref{togen}) and (\ref 
{toback})), we have  
\begin{equation} 
\Phi^{*}({\bf s};\xi )=
\int_{\mbox{{\scriptsize \boldmath$\kappa$}}}
\sum_{{\bf s_0}}^{(-)}
e^{-i({{\bf s}}-{\bf s_0})
\mbox{{\scriptsize \boldmath$\kappa$}} }
\tilde{\Phi}({\bf 0},\mbox{\boldmath$\kappa$}|-
{\bf s_0};\xi ,1)\;\;\;.\label{sumgf} 
\end{equation} 
where the sums over $\nu$ and ${\bf r}$
 have been carried out and the 
explicit dependence on ${\bf s_0}$
 restored. Referring to 
eqns. (\ref{elso})-(\ref{explGammaP}), we have  
\begin{equation} 
\tilde{\Phi}({\bf 0},\mbox{\boldmath$\kappa$}|-
{\bf s_0};\xi ,1)=\frac{1}{1-\xi }\left\{ 
1+\xi \sum_{\{{\bf a},{\bf b}\}} 
\overline{\Gamma }({\bf 0},
\mbox{\boldmath$\kappa$},{\bf a};1)\overline{L}_{ 
{\bf a},{\bf b}}(\mbox{\boldmath$\kappa$};
\xi ,1)P^F({\bf b}|-{\bf  
s_0};\xi )\right\} \;,  \label{PHIINT} 
\end{equation} 
\begin{equation} 
\overline{\Gamma }({\bf 0},
\mbox{\boldmath$\kappa$},{\bf a};1)=p\left( e^{i 
\mbox{\boldmath$\kappa$}{\bf \cdot }
{\bf a}}-1\right)   \label{GammaINT} 
\end{equation} 
and  
\begin{equation} 
(\overline{{\bf L}}^{-1}(\mbox{\boldmath$\kappa$};
\xi ,1))_{{\bf a},{\bf b} 
}=\delta _{{\bf a},{\bf b}}-\xi p\left\{ e^{i 
\mbox{{\scriptsize 
\boldmath$\kappa$}}\cdot {\bf b}}
P^F({\bf a}|-{\bf b};\xi 
)-u\right\} \quad .  \label{LinvINT} 
\end{equation} 
 
At this point, we will restrict our 
attention to $d=2$ only. Clearly, there 
is no interesting behavior in  $d=1$. 
On the other hand, the random walk 
being transient in  $d>2,$ disorder 
will be confined in the large time 
limit.  
 
In two dimensions, the matrix (\ref{LinvINT})
 is $4\times 4$ and must be 
inverted laboriously. The calculations 
involved from (\ref{sumgf}) to (\ref 
{LinvINT}) are reasonably straightforward,
 but extremely lengthy. We shall 
present only the final result obtained
 in the $n\gg 1$ limit. 
 
For {\em fixed} ${\bf s}=(s_1,s_2)$, we obtain  
\begin{equation} 
\phi _n^{*}({\bf s})=\frac{1}{2}-
\frac{1+2s_2}2\;\sqrt{\pi (\pi -1)}\;\frac{1}{ 
\sqrt{\ln {n}}}-\frac{\ln {\ln {n}}}{\ln {n}}+
{\cal O}\left( \frac 1{\ln {n} }\right) \;. 
\end{equation} 
It is clear that the complete disorder
 is the final state, given by $\phi 
^{*}=1/2$, i.e., ``complete gray''. As
 expected, this value is approached 
from below if $s_2>0$, and from above
 if $s_2<0$. Since the frequency of 
``kicks'' scales as $\sqrt{\ln {n}}$, we see that the 
decay follows $1/\sqrt{ 
\ln {n}}$ rather than the typical decay 
of a random walk, i.e., $1/\sqrt{n}$. 
 
Instead of fixed ${\bf s}$, we seek 
a scaled distribution. 
For this we note first that disorder 
{\em along} the interface should arise 
relatively quickly, since it depends 
only on the presence of the random 
walker, which wanders afar as $\sqrt{n}$.
 On the other hand, disorder in the 
vertical direction (from the origin) 
relies entirely on the wandering of the 
passive walkers, so that it will occur 
at the $\sqrt{\ln {n}}$ time scale. 
The appropriate scaling variables 
turn out to be  
\[ 
x=2s_1/\sqrt{n}\quad \mbox{ and}\quad
 y=2s_2\sqrt{\pi (\pi -1)/\ln {n}} 
\] 
while the final result is  
\begin{equation} 
\left[ \phi _n^{*}(x,y)-\Theta (-y)\right] \ln {8n}
=sgn(y)K_0(|x|)e^{-|y|}+ 
{\cal O}\left( \frac{1}
{\sqrt{\ln {n}}}\right)   \label{last} 
\end{equation} 
where $sgn(y)$ denotes the sign of $y$ 
and $K_0$ is the modified Bessel 
function. Note that the square bracket
 on the left hand side is a measure of 
the disorder, since $\Theta $ is the 
step function that represents the 
initial probability distribution for 
finding black particles. Finally, since  
$K_0(z)\rightarrow \sqrt{\pi /2z}\exp {(-z)}$
 for large $z$, we see that the 
decay in both directions are dominated
 by exponentials (in the scaling 
variables $x$ and $y$).  
 
\section{Summary and Outlook} 
 
We have presented a study of a class of 
random walks on a $d$ dimensional 
hypercubic lattice in which the walker 
hops from site to nearest neighbor 
site with one exception. It {\em hops 
over} a particular site. In case the 
hop-over rates are isotropic, the full 
probability distribution was found 
explicitly. We have also investigated 
the statistics associated with the 
hop-overs, so that we can find both how 
often and in which direction the 
hop-overs occur. The latter study can 
be readily applied to the behavior of 
a tagged particle, which moves only if 
a mobile vacancy were to exchange 
places with it. All previously known 
results of tagged diffusion are easily 
recovered in this simpler, novel approach.
 Finally, we showed how this study 
can be applied to two physical examples 
of the vacancy mediated disordering 
process. 
 
Though we have focussed only on infinite 
systems, our methods are readily 
generalizable to finite (periodic) lattices.
 In three or higher dimensions, 
such a generalization is crucial, since the 
random walk is transient, so 
that an infinitesimal density of vacancies
 can not give rise to system-wide 
dynamics. One way to estimate the effects 
of finite density is to consider 
finite systems with periodic boundary 
conditions. Another interesting 
generalization is the {\em biased} random 
walk. With a bias, there would be 
little reason to study hop-over walks on 
an infinite lattice, since the 
walker never returns to the hop-over site. 
By contrast, on a {\em finite} 
lattice, the walker will ``run into'' the
 special site periodically. In the 
steady state, i.e., for times large 
compared to the traverse time, we can 
expect an asymmetric distribution around 
the ``defect''. If the hop-over 
rates are also biased, say, in the opposite
 direction, it may be possible 
for an effective binding to occur. This 
problem can be mapped into a 
limiting version of the biased diffusion 
of two species introduced sometime 
ago \cite{SHZ}. It would be very interesting 
to examine the walker-defect 
distributions and check if long range 
correlations \cite{KSZ} also appear. 
 
\section{Acknowledgments} 
 
Illuminating discussions with 
G. Korniss, B. Schmittmann and 
W. Triampo are gratefully 
acknowledged.  This research is supported in part by 
the US National Science Foundation through
 the Division of Material Research 
and the Hungarian Science Foundation under
 grant numbers OTKA F17166,
T17493 and T19483.

\end{document}